# A Multi-Modal Deep Learning Framework for Pan-Cancer Prognosis

Binyu Zhang[a,*], Shichao Li[a,*], Junpeng Jian[a], Zhu Meng[a,b], Limei Guo[c], Zhicheng Zhao[a,b,**]

[a]School of Artificial Intelligence, Beijing University of Posts and Telecommunications, Beijing, China
[b]Beijing Key Laboratory of Network System and Network Culture, Beijing, China
[c]Department of Pathology, School of Basic Medical Sciences, Third Hospital, Peking University, Beijing, China

[*]These authors contributed equally to this work and considered as co-first author.
[**]Corresponding author.

E-mail address: zhangbinyu@bupt.edu.cn (B. Zhang), lishichao@bupt.edu.cn (S. Li), 2521jian@bupt.edu.cn (J. Jian), bamboo@bupt.edu.cn (Z. Meng), guolimei@bjmu.edu.cn (L. Guo), zhaozc@bupt.edu.cn (Z. Zhao)

## Summary

**Background**

Prognostic task is of great importance as it closely related to the survival analysis of patients, the optimization of treatment plans and the allocation of resources. The existing prognostic models have shown promising results on specific datasets, but there are limitations in two aspects. On the one hand, they merely explore certain types of modal data, such as patient histopathology whole slide images (WSI) and gene expression analysis, while neglecting information related to patients' basic characteristics, medical records and treatment regimens. On the other hand, they adopt the per-cancer-per-model paradigm, which means the trained models can only predict the prognostic effect of a single type of cancer, resulting in weak generalization ability.

**Methods**

In this paper, a deep-learning based model, named Unified Multi-modal Pan-cancer Survival Network (UMPSNet), is proposed. Specifically, to comprehensively understand the condition of patients, in addition to constructing encoders for histopathology images and genomic expression profiles respectively, UMPSNet further integrates four types of important meta data (demographic information, cancer type information, treatment protocols, and diagnosis results) into text templates, and then introduces a text encoder based on finetuned CLIP[20] to extract textual features. In addition, the optimal transport (OT)-based attention mechanism is utilized to align and fuse features of different modalities. Furthermore, a guided soft mixture of experts (GMoE) mechanism is introduced to effectively address the issue of distribution differences among multiple cancer datasets. The code of UMPSNet is available at https://github.com/binging512/UMPSNet.

**Findings**

UMPSNet conducts joint training and validating for survival prediction with five public multi-type cancer datasets from The Cancer Genome Atlas (TCGA, which contains 2831 cases and a total of 3523 WSIs), including Bladder Urothelial Carcinoma (BLCA, 373 cases), Breast Invasive Carcinoma (BRCA, 956 cases), Glioblastoma & Lower Grade Glioma (GBMLGG, 569 cases), Lung Adenocarcinoma (LUAD, 453 cases), and Uterine Corpus Endometrial Carcinoma (UCEC, 480 cases). UMPSNet achieves an average concordance index (C-index) of 0.725 on the five datasets and outperforms all state-of-the-art (SOTA) methods. This result not only demonstrates the effectiveness of our UMPSNet, but also shows the feasibility and great potential of the new learning paradigm, that is, carrying out collective prognostic analysis for multiple diseases.

**Interpretation**

By incorporating the multi-modality of patient data and joint training, a deep-learning based approach, named UMPSNet is proposed to address the problem of pan-cancer prognosis. UMPSNet outperforms all SOTA approaches, and moreover, it demonstrates the effectiveness and generalization ability of the proposed learning paradigm of a single model for multiple cancer types. This advancement highlights the potential to enhance clinical decision-making and promote precision medicine. It also provides a new solution and a more comprehensive and effective tool for personalized patient care.

**Funding**

This work is supported by Chinese National Natural Science Foundation (62401069), and BUPT Excellent Ph.D. Students Foundation (CX2022150).

**Introduction**

Deep learning approaches have revolutionized numerous medical applications in contemporary healthcare systems[1,2,3]. Patient prognosis prediction, as one of the fundamental tasks in clinical medicine, has attracted substantial interest[4]. By systematically evaluating disease progression and patient condition, healthcare providers can generate survival estimates, facilitating optimal treatment and personalized therapeutic interventions.

Existing deep learning-based prognostic methods rely predominantly on histopathological WSIs, which represent the gold standard for cancer staging diagnosis and demonstrate strong correlation with patient survival outcomes[5,6]. However, constrained by the ultra-high resolution of WSI and Graphic Process Unit (GPU) memory, researchers are forced to adopt multi-instance learning (MIL) methods to train the network, where pre-trained models are employed to extract features from WSI patches for subsequent training. Nevertheless, the regions that are most critical to patient survival may occupy only a small portion of the WSI. Therefore, the genomic expression profiles are introduced to quantify the severity of cancer[7,8,9]. Additionally, survival outcomes are influenced by multiple factors beyond disease severity, patients' physical condition, psychological state, cancer type, and therapeutic interventions will all have an impact. However, these factors are overlooked by the existing methods, consequently, they can only achieve suboptimal performance.

Moreover, different types of cancers have numerous differences[10], such as the distribution of histopathological image features, genomic expression profiles, and survival times, etc. Consequently, existing approaches are restricted to training and evaluation on single-cancer datasets, which leads to low generalization ability.

Therefore, in this paper, aiming to break through the limitation of the per-cancer-per-model paradigm, we propose UMPSNet, a new and unified multi-modal network for pan-cancer prognosis. Firstly, in order to enable UMPSNet to conduct learning as comprehensively as possible, six aspects of information about patients are collected and encoded: patients' histopathological WSIs, genomic expression profiles, demographic information, cancer type information, treatment protocols, and diagnosis results. Secondly, considering the differences in modalities and information, UMPSNet employs different feature representation strategies. For example, histopathological WSIs are split into patches with the same sizes and encoded via a pre-trained visual model, while genomic expression profiles are separated into groups according to their biological roles. Regarding the meta data, as they are usually discrete numerical values, specific text templates are firstly designed to incorporate them, and then a text encoder is utilized to extract the features. Thirdly, an OT-based attention mechanism[9] is leveraged to align and fuse features of multiple modalities. Finally, a survival classifier is trained to predict the patient survival risk, while a cancer type classifier is also learned to predict the cancer type as an agent task. Moreover, inspired by the MoE[11,12], we introduce a GMoE mechanism to fuse all features, which is capable of identifying common characteristics of different cancer types while adaptively extracting the features specific to each cancer type.

Our contribution can be summarized as follows:

- We propose a unified multi-modal pan-cancer prognosis model named UMPSNet, where six types of complementary patient data, including images, genes and texts, are firstly encoded by three different modal encoders, respectively. Then, OT algorithm is embedded into the attention mechanism to extract inter-related features among modalities. In addition, a cancer type classifier is learned as an agent task to improve the model capability of distinguishing different cancers.
- To address the distributional differences among various cancer types, UMPSNet introduces the GMoE mechanism, which employs multiple experts to collaboratively accomplish the prognosis task. Moreover, to learn common knowledge of different cancer types while preserving individualized information for each patient, the text information is utilized to guide the weighted fusion of the outputs from different expert modules.
- To the best of our knowledge, UMPSNet is the first model designed for pan-cancer prognosis via a joint learning paradigm. Extensive experiments on five public datasets in TCGA, namely BLCA, BRCA, GBMLGG, LUAD and UCEC, demonstrate that UMPSNet outperforms all state-of-the-art pan-cancer prognosis methods, while achieves competitive performance with those for single-cancer. These results not only validate the effectiveness of UMPSNet in diverse cancer types, but also show its potential to become a new benchmark for joint training of pan-cancer prognosis.

## Methods

**Problem formulation**

Due to ultra-high resolution of Whole Slide Images (WSIs), the common MIL scheme is adopted, and the multi-modal data is first preprocessed to form data bags applicable to model training.

**WSI bag formulation.** Considering that the slides of patients contain a large amount of background, thereby a foreground segmentation algorithm is firstly employed, and then the foreground regions are partitioned into uniform-sized patches as the WSI data bag:

$$X_i^p = \{x_{i,n}^p\}_{n=1}^{N_{p,i}}, \tag{1}$$

where $x_{i,n}^p$ is the $n$th patch for $i$th patient, and $N_{p,i}$ is the total number of patches that belong to the patient.

**Genomic bag formulation.** Given that the genes have different functions, we categorize them into six groups: (i) tumor suppressor genes, (ii) oncogenes, (iii) protein kinases, (iv) cell differentiation markers, (v) transcription factors, and (vi) cytokines and growth factors. To normalize the genomic data, the z-scores relevant to diploid samples are calculated to represent the genomic data bag:

$$X_i^g = \{x_{i,n}^g\}_{n=1}^{N_g} \tag{2}$$

where $x_{i,n}^g$ stands for the $n$th genomic group of the $i$th patient, and $N_g = 6$ represents the number of the genomic groups.

**Text bag formulation.** Inspired by the large language models (LLM)[13][14] and in order to handle complex discrete four types of text data[11], four text templates are designed to standardize each piece of the original text data. Specifically, for the patient demographic, including sex, age and race, a simple template is designed as: "{He/She} is a {Age}-year-old {Race} race {Man/Woman}". For the cancer type, the template "This is a patient who has {Cancer}.", will be filled with the full name of the cancer. Given that the diagnostic information includes both the overall cancer staging and the TNM staging, the entire template is designed as follows: "{He/She} has {Primary_diagnosis} at {Stage}. {T_stage}, {N_stage}, {M_stage}.". In this template, "Primary_diagnosis" denotes the nature of the cancer, "Stage" represents the overall staging phase, while the subsequent "T/N/M_stage" components are filled with the corresponding text descriptions. Finally, as treatment protocols represent whether radiation or pharmaceutical therapy have been utilized, "{Treatments} is applied." is adopted as the template. To the end, the text bag can be formulated as:

$$X_i^t = \{x_{i,dem}^t, x_{i,can}^t, x_{i,dia}^t, x_{i,tre}^t\}, \tag{3}$$

where $x_{i,dem}^t$, $x_{i,can}^t$, $x_{i,dia}^t$, and $x_{i,tre}^t$ represent the generated text for patient demographic, cancer type, diagnosis and treatment protocol, respectively.

## Network

To conduct a comprehensive analysis of multi-modal data and accomplish the patient prognosis, UMPSNet is proposed. As shown in Fig.1, it consists of three components: feature extraction, feature interaction, and the classification.

**Feature extraction.** To align and embed the three different modal data, three different encoders are utilized.

*1) WSI data bag*. A pretrained image encoder CTransPath is leveraged to extract the patch features $f_i^p$,

$$f_i^p = \{F^p(x_{i,n}^p)\}_{n=1}^{N_i^p}, \tag{4}$$

where $F^p(\cdot)$ represents the pretrained patch feature extractor.

*2) Genomic data*. Because there are differences and missing data in genomic data among cancer datasets, we apply zero-padding to fill in the missing data and generate a mask simultaneously. In the mask, the positions that are not zero-padded are designated as 1, while the others are set as 0. Then, Transformers[15] with position embedding are utilized to extract the genomic features $f_i^g$, which can be written as,

$$f_i^g = \{F_n^g(x_{i,n}^g + PE, M_{i,n}^g)\}_{n=1}^{N^g}, \tag{5}$$

where $F_n^g(\cdot)$ is the Transformer encoder for the $n$th group of genomic data, $M_{i,n}^g$ stands for the corresponding generated attention mask, and $PE$ represents the position embedding.

$$\begin{cases} PE_{(pos,2m)} = \sin\left(\frac{pos}{10000^{\frac{2m}{d_{model}}}}\right) \\ PE_{(pos,2m+1)} = \cos\left(\frac{pos}{10000^{\frac{2m}{d_{model}}}}\right) \end{cases}, \tag{6}$$

where $pos$ represents the position of the gene in the genomic data, $d_{model}$ is the dimension of the embedding features. Experiments show that the above-mentioned method effectively relieves the issue of missing data while avoiding the inaccuracies caused by zero-padding.

*3) Text data.* An LLM is leveraged to extract the text features. Specifically, we finetune the CLIP[20] with adapters to enable the LLM to retain its generalization capability while adapting to UMPSNet. The text features $f_i^t$ can be written as,

$$f_i^t = \{F^t(x_{i,n}^t)\}_{n=1}^{N^t}, \tag{7}$$

where $F^t(\cdot)$ is the adapter-embedded CLIP model, $N^t = 4$ is the number of text types.

**Feature Interaction.** In order to alleviate the negative impacts of modal differences on model learning, we firstly align all features into a unified latent space by means of OT-based attention mechanism. Note that OT-based attention is applied between image features and text features, as well as between genomic features and text features, thus two optimal matching flows $A_i^X$ can be obtained.

$$A_i^X = F_{OT}^X(f_i^X, f_i^t), \tag{8}$$

where $F_{OT}^X(\cdot)$ is the OT-based attention module, and $X$ can be either $p$ or $g$, representing the image branch or the genomic branch.

Secondly, the features from different modalities can be aligned via the matrix multiplication with the matching flow matrix. OT-based attention not only aligns different features, but also reduces the dimension of features, especially that of image features. Afterwards, text-guided Transformer layers are adopted to extract patient-aware features from each modality. Specifically, considering that the text features contain a wealth of patient-specific information, they are treated as Query in the Transformer decoder structure, while the image or genomic features are regarded as Key and Value. Finally, the fused image features $f_i^{p'}$, genomic feature $f_i^{g'}$, and the original text features $f_i^t$ are fed into the classification module in a combined manner.

**Classification.** In this module, GMoE mechanism is firstly proposed to re-fuse the features and predict hazard scores of the patient. In addition, to enhance the model's capacity to distinguish among various cancer types, we design an agent task as an auxiliary component to jointly train the network, that is, a dedicated classifier is trained to predict the possibility of the cancer types. Note that text features are not utilized in the agent task, since they inherently contain information about cancer types.

## GMoE architecture

As an expansible soft MoE, the GMoE is constructed to reduce the data distribution discrepancies within pan-cancer datasets. It consists of multiple expert modules, as shown in Fig. 2(a), each one has an identical structure, yet their parameters are not shared.

The architecture of each expert is shown in Fig. 2(b), where two Transformer decoder layers are adopted to fuse the features. Additionally, we utilize the text features as queries to guide the module to obtain the embedding related to patient survival from the image and genomic features. Finally, the embedding is fed into a survival classifier to predict the survival scores. Based on our GMoE architecture, the features of different cancer types will activate different expert modules, facilitating predictions as well as providing reasonable explanations.

The above prediction process can be formulated as follows as

$$S_{ha,i} = \{F_n^e(f_i^{p'}, f_i^{g'}, f_i^t)\}_{n=1}^{N_e} \times F^{cw}(x_i^{can}, x_i^{dia}) \tag{9}$$

where $S_{haz,i}$ represents the predicted hazard score of the $i$th patient, $F_n^e(\cdot)$ is the $n$th expert in GMoE architecture, while $N_e$ stands for the total number of experts, $F^{cw}(\cdot)$ denotes the linear layer for generating weights, and $x_i^{can}$ and $x_i^{dia}$ are the cancer type and diagnosis of the patient, respectively.

Finally, the model obtains the survival prediction results by performing a weighted summation of the outputs from all experts. The weights are generated based on the patient's cancer type and diagnostic results, thus enhancing the model's ability to distinguish among different cancer types.

## Loss functions

To train the whole model, two loss functions are considered. Firstly, the negative log-likelihood loss ($L_{nll}$)[8] is introduced to supervise the survival prediction task,

$$L_{nll}(S_{haz,i}, c_i, y_i) = -c_i \cdot \log(S_{surv,i}(y_i)) - (1 - c_i) \cdot \log(S_{surv,i}(y_i - 1)) - (1 - c_i) \cdot (S_{ha,i}(y_i)), \tag{10}$$

where $c_i$ denotes the censorship of the $i$th patient, $S_{haz,i}(y_i)$ stands for the hazard score at the $y_i$ time bin. $S_{surv,i}(y_i)$ represents the cumulative survival probability at the true survival time bin $y_i$ and can be expressed as,

$$S_{surv,i}(y_i) = \prod_{n=0}^{y_i}(1 - S_{haz,i,n}), \tag{11}$$

where $S_{haz,i,n}$ is the predicted hazard score for the $i$th patient at $n$th time bin.

Secondly, the cross-entropy loss ($L_{ce}$) is adopted to supervise the classification task for five cancer types.

Overall, the total loss function can be written as,

$$L_{total} = L_{ce}(\hat{y}_{can,i}, y_{can,i}) + L_{nll}(S_{haz,i}, c_i, y_{surv,i}), \tag{12}$$

where $L_{nll}(\cdot)$ and $L_{ce}(\cdot)$ represent the NLLLoss and CELoss, respectively, $y_{surv,i}$ stands for the ground truth of the patient survival bin, while $\hat{y}_{can,i}$ and $y_{can,i}$ denote the predicted score and the actual cancer type of the patient, respectively.

## Results

### Datasets and evaluation metrics

To evaluate the performance of UMPSNet, five public cancer datasets in TCGA including BLCA, BRCA, GBMLGG, LUAD, UCEC, are leveraged, and their statistical information is shown in Table 1. TCGA is a large-scale, multi-institutional research initiative jointly launched by the National Cancer Institute (NCI) and the National Human Genome Research Institute (NHGRI) in the United States. This project involves the systematic collection of extensive data from numerous leading cancer research institutions and hospitals nationwide. To evaluate the model performance on pan-cancer prognosis, these five datasets are merged as a joint training dataset. Following MOTCat[9], a five-fold validation process is adopted. The concordance index

(C-index)[16] is selected as the metric. The C-index metric pairs patients and calculates the proportion of all pairs that are consistent with the relationship between the actual risk and predicted one.

Table 1. The details of the datasets. For the genomics, TSG, ONC, PK, CDM, TF and CGF represent Tumor Suppressor Genes, Oncogenes, Protein Kinases, Cell Differentiation Markers, Transcription Factors and Cytokines & Growth Factors respectively.

| Datasets | BLCA | BRCA | GBMLGG | LUAD | UCEC |
|---|---|---|---|---|---|
| Cases | 373 | 956 | 569 | 453 | 480 |
| Slides | 437 | 1017 | 1014 | 516 | 539 |
| Gender | | | | | |
|   Male | 277 | 11 | 330 | 208 | 0 |
|   Female | 96 | 945 | 239 | 245 | 480 |
| Age (years, mean, std) | 68.0(10.6) | 58.2(13.1) | 45.9(14.4) | 65.1(9.9) | 63.9(11.1) |
| Race | | | | | |
|   White | 297 | 663 | 522 | 343 | 325 |
|   Black or African American | 20 | 155 | 26 | 50 | 96 |
|   Asian | 42 | 57 | 9 | 7 | 19 |
|   American Indian or Alaska native | 0 | 1 | 1 | 1 | 3 |
|   Native Hawaiian or other pacific islander | 0 | 0 | 0 | 0 | 8 |
|   Not reported | 14 | 80 | 11 | 52 | 29 |
| Genomic | | | | | |
|   TSG | | | 82 | | |
|   ONC | | | 328 | | |
|   PK | | | 513 | | |
|   CDM | | | 443 | | |
|   TF | | | 1536 | | |
|   CGF | | | 452 | | |
| Treatment | | | | | |
|   None | 223 | 167 | 146 | 260 | 183 |
|   Radiation | 25 | 38 | 76 | 29 | 110 |
|   Pharmaceutical | 99 | 292 | 52 | 102 | 59 |
|   Both | 26 | 459 | 295 | 62 | 128 |
| Survival (uncensored, censored) | | | | | |
|   < 12 months | 89(34) | 18(131) | 142(173) | 50(41) | 19(46) |
|   12~24 months | 63(80) | 19(304) | 82(202) | 53(129) | 22(130) |
|   24~36 months | 26(45) | 22(122) | 39(102) | 55(54) | 21(91) |
|   36~60 months | 16(32) | 28(145) | 33(132) | 35(50) | 10(85) |
|   >=60 months | 7(45) | 47(185) | 25(84) | 9(40) | 7(108) |

**Implementation details**

In the process of WSI data bag formulation, OTSU algorithm[17] is applied to segment the tissue regions, then the regions are split into non-overlapping patches, with a resolution of 256×256 pixels under a 20× magnification[18]. As for the image encoder, self-trained CTransPath[19] is adopted to extract the WSI patch features for training. Note that, CTransPath specifically leverages histopathological images in the self-training stage, thus can extract more discriminative deep features. In the process of genomic data bag formulation, with the attention mask, Transformer[15] is employed to encode each group of genomic data into genomic features. For the text bag, four types of text prompts are encoded via CLIP[20] with adapter layers[21,22]. Note that only the adapter layers are fine-tuned during model training while CLIP model remains frozen.

In the model training, AdamW optimizer is adopted, with an initial learning rate of $2 \times 10^{-4}$ and a weight decay of $1 \times 10^{-5}$. Due to the different number of patches for each patient, the batch size is set as 1, while the gradient update step is set as 32 to accumulate the gradient. All the experiments are conducted on a server with the Ubuntu 18.04 LTS OS, an Intel CPU E5 2.20GHz CPU and one NVIDIA Tesla V100 GPU for training and testing. Python version is 3.10.14, CUDA version is 10.2 and PyTorch version is 1.12.1.

**Experimental results**

**Comparison with the SOTA methods**. Following the dataset partition in MOTCat[9], we conduct 5-fold validation experiments on the joint dataset. As shown in Table 2, comparing with existing approaches, UMPSNet not only achieves new state-of-the-art performance among the joint training methods, but also achieves competitive performance against the separately training approaches.

Table 2. The comparison of UMPSNet with existing approaches on five TCGA datasets. ST represents separate training the five datasets. P, G, and T denote histopathological WSI, genomic data and text input, respectively. * represents that the metric is obtained via reproducing the related work. The best results are highlighted in **bold** respectively.

| Train | Methods | P | G | T | Datasets | | | | | Overall |
|---|---|---|---|---|---|---|---|---|---|---|
| | | | | | BLCA | BRCA | GBMLGG | LUAD | UCEC | |
| ST | SNN[7] | | √ | | 0.618±0.022 | 0.624±0.060 | 0.834±0.012 | 0.611±0.047 | 0.679±0.040 | 0.673 |
| | SNNTrans[7,23] | | √ | | 0.659±0.032 | 0.647±0.063 | 0.839±0.014 | 0.638±0.022 | 0.656±0.038 | 0.688 |
| | AttnMIL[24] | √ | | | 0.551±0.049 | 0.577±0.043 | 0.786±0.026 | 0.561±0.078 | 0.639±0.057 | 0.623 |
| | DeepAttnMISL[25] | √ | | | 0.504±0.042 | 0.524±0.043 | 0.734±0.029 | 0.548±0.050 | 0.597±0.059 | 0.581 |
| | CLAM-SB[18] | √ | | | 0.559±0.034 | 0.573±0.044 | 0.779±0.031 | 0.594±0.063 | 0.644±0.061 | 0.630 |
| | CLAM-MB[18] | √ | | | 0.565±0.027 | 0.578±0.032 | 0.776±0.034 | 0.582±0.072 | 0.609±0.082 | 0.622 |
| | TransMIL[23] | √ | | | 0.608±0.139 | 0.626±0.042 | 0.798±0.033 | 0.641±0.033 | 0.657±0.044 | 0.666 |
| | DTFD-MIL[26] | √ | | | 0.546±0.021 | 0.609±0.059 | 0.792±0.023 | 0.585±0.066 | 0.656±0.045 | 0.638 |
| | Surformer[27] | √ | | | 0.594±0.027 | 0.628±0.037 | 0.809±0.026 | 0.591±0.064 | 0.681±0.028 | 0.661 |
| | G-HANet[28] | √ | | | 0.630±0.032 | 0.664±0.065 | 0.817±0.022 | 0.612±0.028 | **0.729±0.050** | 0.690 |
| | Propoise[29] | √ | √ | | 0.636±0.024 | 0.652±0.042 | 0.834±0.017 | 0.647±0.031 | 0.695±0.032 | 0.693 |
| | MCAT[8] | √ | √ | | **0.672±0.032** | 0.659±0.031 | 0.835±0.024 | 0.659±0.027 | 0.649±0.043 | 0.695 |
| | SurvPath[30] * | √ | √ | | 0.614±0.061 | 0.656±0.036 | 0.795±0.025 | 0.620±0.036 | 0.691±0.024 | 0.675 |
| | MOTCat[9] * | √ | √ | | 0.664±0.031 | 0.658±0.012 | 0.839±0.027 | 0.668±0.036 | 0.670±0.038 | 0.699 |
| | MOTCat[9] *+Text | √ | √ | √ | 0.671±0.019 | **0.681±0.018** | 0.847±0.017 | 0.700±0.029 | 0.682±0.016 | 0.716 |
| Joint | MCAT[8] * | √ | √ | | 0.583±0.053 | 0.564±0.012 | 0.798±0.018 | 0.640±0.051 | 0.645±0.030 | 0.646 |
| | MOTCat[9] * | √ | √ | | 0.573±0.018 | 0.576±0.050 | 0.812±0.007 | 0.646±0.053 | 0.633±0.042 | 0.648 |
| | MOTCat[9] *+Text | √ | √ | √ | 0.601±0.028 | 0.614±0.037 | 0.840±0.026 | 0.672±0.063 | 0.640±0.058 | 0.674 |
| | UMPSNet | √ | √ | √ | **0.659±0.038** | **0.732±0.046** | 0.839±0.012 | 0.662±0.016 | **0.730±0.051** | **0.725** |

UMPSNet achieves an average C-index score of 0.725 on the five datasets, and the scores on the BLCA, BRCA, GBMLGG, LUAD and UCEC datasets are 0.659, 0.732, 0.839, 0.662 and 0.730 respectively. It outperforms MOTCat by 3.7% and 11.9% in separate and joint training manner, respectively.

To make a fair comparison, we also integrate the same text encoder and text input into MOTCat by concatenating other features, and denote it as MOTCat+text. The results show that, in the separate training mode, MOTCat+text surpasses other approaches, which demonstrates that the model ability can be improved via supplementing patient' textual information. Moreover, UMPSNet also outperforms MOTCat+Text in joint training manner, which indicates that the proposed text-guided structure, the agent task, and the GMoE mechanism enable the model to better adapt to multiple cancer types. Besides, it demonstrates that during the training, UMPSNet has not only learned the common features that affect the survival periods of patients with different cancers, but also retained the unique characteristics specific to each cancer type.

To better illustrate the statistical differences in performance between the UMPSNet and other approaches, the predictive results of each model are visualized via Kaplan-Meier (KM) curves in which the lower the P-value is, the better the model performance will be, as shown in Fig.3. Specifically, based on the patient risk scores predicted by the models, all patients are divided into high-risk and low-risk groups on average. Subsequently, survival curves corresponding to the ground truth survival times of each patient are plotted. Additionally, a Logrank test is conducted to measure the statistical significance between the two patient groups. It can also be observed that UMPSNet exhibits a significant gap between the low-risk and high-risk lines, achieving lower P-values on most of datasets.

**Ablation Study**

**The effectiveness of each module**. To further explore the effectiveness of each module within UMPSNet, the ablation experiments are conducted. As illustrated in Table 3, the performance of UMPSNet progressively improves with the incremental incorporation of various modules into the network.

Table 3. The ablation study of UMPSNet with different modules. The best results are highlighted in **bold**.

| Method | GMoE | Agent | Datasets | | | | | Overall |
|---|---|---|---|---|---|---|---|---|
| | | | BLCA | BRCA | GBMLGG | LUAD | UCEC | |
| MOTCat[9]* | | | 0.573±0.018 | 0.576±0.050 | 0.812±0.007 | 0.646±0.053 | 0.633±0.042 | 0.648 |
| UMPSNet | | | 0.634±0.052 | 0.689±0.074 | 0.823±0.027 | 0.641±0.054 | 0.680±0.059 | 0.694 |
| | √ | | 0.658±0.048 | **0.740±0.037** | 0.834±0.019 | 0.645±0.048 | 0.692±0.064 | 0.714 |
| | √ | √ | **0.659±0.038** | 0.732±0.046 | **0.839±0.012** | **0.662±0.016** | **0.730±0.051** | **0.725** |

These experimental results underscore the necessity of incorporating diverse modality data to enhance the performance of patient survival predictions, and also demonstrate the effectiveness of the network modules and training strategies designed by us. In addition, the two methods we proposed, that is, using text as queries to guide feature extraction and taking cancer classification as a supervision task to train the image branch and the text branch, both contribute to the extraction of more discriminative features, thereby improving its performance in pan-cancer survival prediction.

**The number of experts**. To determine the optimal number of experts, we also carry out a series of experiments. As shown in Table 4, UMPSNet shows the best performance when $N_e$ is set to 10. It demonstrates that the insufficient experts cannot meet the requirements for distinguishing multiple cancers, and meanwhile, an excessive number of experts will lead to difficulties in training optimization.

Table 4. The ablation study of UMPSNet with different number of experts in MoE module. The best results are highlighted in **bold**.

| $N_e$ | Datasets | | | | | Overall |
|---|---|---|---|---|---|---|
| | BLCA | BRCA | GBMLGG | LUAD | UCEC | |
| 1 | 0.644±0.044 | 0.703±0.041 | 0.828±0.023 | 0.680±0.057 | 0.699±0.060 | 0.711 |
| 5 | 0.641±0.042 | **0.738±0.025** | 0.840±0.025 | **0.694±0.025** | 0.703±0.036 | 0.723 |
| 10 | **0.659±0.038** | 0.732±0.046 | 0.839±0.012 | 0.662±0.016 | **0.730±0.051** | **0.725** |
| 15 | 0.634±0.032 | 0.731±0.011 | **0.847±0.011** | 0.644±0.039 | 0.687±0.049 | 0.709 |
| 20 | 0.641±0.050 | 0.731±0.054 | 0.827±0.022 | 0.666±0.012 | 0.642±0.069 | 0.701 |

**Interpretability**

To demonstrate the interpretability of UMPSNet, GradCAM[31] is utilized to generate class activation maps (CAMs) for the input images. As illustrated in Fig.4, the red regions depict areas that pose greater threats to patient survival, while the blue regions are the opposite.

Additionally, CAMs for genomic data are also generated to analyze the impact of differential gene expression on various types of cancer. Through averaging the gene CAMs of all sample for each cancer dataset, we initially identify the top 3 genes that have the most significant impact on patient survival in each genomic group, as shown in Fig.5. This shows the interpretability of our model and also provides important references for subsequent medical practices and research.

**Discussion**

Prognosis aims to predict the survival risk of the patients so as to optimize the treatment plans and the resource allocation etc. Thus, a number of prognosis approaches are proposed. As the gold standard for cancer diagnosis, histopathological WSIs are widely utilized in this task. DeepAttnMISL[25] proposes siamese MI-FCN to learn features from different phenotype cluster and introduces attention-based MIL pooling to perform a trainable weighted aggregation. To build hierarchical representations of morphological image patch features, Patch-GCN[32] treats WSIs as 2D point clouds and formulates a graph-based structure, then leverages graph convolutional network (GCN) to aggregate the information from the patches. Surformer[27] introduces a RRCA module to simultaneously detect global features and multiple pattern-specific local features, and proposes a disentangling loss constraining the local features to focus on distinct patterns. To enhance the model performance, LNPL-MIL[33] and HistMIMT[34] adopt agent tasks to train the model, such as cancer diagnosis, genomic expression prediction, etc. As another key characteristic of cancer, differences in genomic expression can also be leveraged to assess cancer progression. Therefore, SNN[7] designs a module to better accommodate the characteristics of genomic data.

To obtain a more comprehensive analysis for patients, several approaches explore to use multi-modal data for joint prediction of patient survival risk. Among these, the main-stream approaches involve the joint analysis of WSIs and genomic data. Pathomic[35] takes WSI patches, cell spatial graph, and genomic profiles into account, and fuses the features for survival prediction. However, the fusion method lacks intermediate feature integration. Therefore, MCAT[8] proposes a genomic-guided co-attention module, which is capable of integrating both WSI and genomic features. In addition, MOTCat[9] embeds OT algorithm into the attention module and achieves impressive results. Moreover, to analyze the influence of individual genes on patient survival, Survpath[30] introduces a transcriptomics tokenizer to generate biological pathway tokens, which leverage the existing knowledge of cellular biology. Meanwhile, a multi-level interpretability framework is proposed to enable deriving unimodal and cross-modal insights about the prediction for prognosis. To reduce the reliance on genomic data during testing while maintaining high performance, G-HANet[28] trains a network using knowledge distillation technique via reconstructing genomic expression profiles from histopathological image features.

Although the existing methods have achieved impressive results in prognosis task, the factors they take into account are still limited. Prognosis task differs from diagnostic task. Multiple factors, including patient's individual condition, psychological state, treatment protocols, and the healthcare environment etc., can all affect survival outcomes. Therefore, UMPSNet aims to learn more valuable features from WSIs, genomic data, and patient' textual information. Additionally, we observe that existing methods usually train and test only on individual datasets, which makes them unsuitable for pan-cancer survival prediction. Meanwhile, recent studies[4] indicate that there are certain commonalities in the survival threats of different cancers. Consequently, UMPSNet attempts to combine multiple cancer types for training to enhance the network's ability to learn their commonalities. In the meantime, UMPSNet will activate different modules within the GMoE mechanism according to the specific cancer type, thus preserving the uniqueness of each cancer.

**Conclusion**

In this study, we introduce UMPSNet, a new multi-modal learning network for pan-cancer prognosis. Specifically, UMPSNet utilizes six kinds of patient data, including histopathological WSI, genomic expression profile, demographic information, cancer type information, treatment protocols, and diagnosis results. Three different encoders are leveraged to extract the features from different modalities, and OT-based attention is applied for feature alignment and interaction. To enhance the model's ability to distinguish between different cancer types, we propose the GMoE mechanism, which utilizes meta data to guide and fuse features. Moreover, a cancer type classifier is also learned as an agent task to improve the model performance. Extensive experimental results demonstrate that UMPSNet not only achieves the state-of-the-art performance in the task of cancer prognosis, but also shows its potential to establish a new benchmark for pan-cancer prognosis.

## Contributors

Binyu Zhang (Conceptualization: Equal; Formal analysis: Lead; Investigation: Equal; Methodology: Lead; Validation: Lead; Visualization: Equal; Writing—original draft: Lead), Shichao Li (Investigation: Equal; Methodology: Equal; Validation: Equal; Visualization: Lead) and Junpeng Jian (Methodology: Equal), Zhu Meng (Conceptualization: Lead, Investigation: Lead), Limei Guo (Conceptualization: Equal; Supervision: Equal) and Zhicheng Zhao (Project administration: Lead, Supervision: Lead, Writing—review & editing: Lead). All authors have full access to all the data, and have final responsibility for the decision to submit for publication.

## Acknowledgments

This work is supported by Chinese National Natural Science Foundation (62401069), and BUPT Excellent Ph.D. Students Foundation (CX2022150).